\documentclass[twocolumn,superscriptaddress,showpacs,preprintnumbers,aps,floatfix]{revtex4}

\usepackage{graphicx}
\DeclareGraphicsExtensions{.jpg,.pdf}
\usepackage{dcolumn}
\usepackage{amssymb,amsmath}
\usepackage{color}
\usepackage{bm}

\begin{document}
\title{Nonlinear dynamical tunneling of optical whispering gallery modes\\ in the presence of a Kerr nonlinearity}
\author{Jeong-Bo Shim}
\affiliation{D\'epartement de Physique, University of Li\`ege, 4000 Li\`ege, Belgium}
\affiliation{Institut f\"ur Theoretische Physik, Otto-von-Guericke-Universit\"at Magdeburg, D-39016 Magdeburg, Germany}
\author{Peter Schlagheck}
\affiliation{D\'epartement de Physique, University of Li\`ege, 4000 Li\`ege, Belgium}
\author{Martina Hentschel}
\affiliation{Institut f\"ur Physik, Technische Universit\"at Illmenau, D-98693, Illmenau, Germany}
\author{Jan Wiersig}
\affiliation{Institut f\"ur Theoretische Physik, Otto-von-Guericke-Universit\"at Magdeburg, D-39016 Magdeburg, Germany}
\begin{abstract}
The effect of a Kerr nonlinearity on dynamical tunneling is studied, using coupled whispering gallery modes in an optical microcavity. The model system that we have chosen is the `add-drop filter', which comprises an optical microcavity and two waveguide coupled to the cavity. Due to the evanescent field’s scattering on the waveguide, the whispering gallery modes in the microcavity form doublets, which manifest themselves as splittings in the spectrum. As these doublets can be regarded as a spectral feature of dynamical tunneling between two different dynamical states with a spatial overlap, the effect of a Kerr nonlinearity on the doublets is numerically investigated in the more general context of the relationship between cubic nonlinearity and dynamical tunneling. Within the numerical realization of the model system, it is observed that the doublets shows a bistable transition in its transmission curve as the Kerr-nonlinearity in the cavity is increased. At the same time, one rotational mode gets dominant over the other one in the transmission, since the two states in the doublet have uneven linewidths. By using coupled mode theory, the underlying mode dynamics of the phenomena is theoretically modelled and clarified.
\end{abstract}
\date{\today}
\maketitle

\section{Introduction}
A whispering gallery mode (WGM) in a microcavity is recognized for its easy experimental realization and high cavity quality factor ($Q$ factor). Using such properties of WGMs, high intensity can be easily induced in a micrometer-scale optics, and the induced high intensity allows for nonlinear optical effects such as the Kerr-nonlinear effect \cite{Boyd2008}, in which the change of refractive index is proportional to the squared value of electric field.

A pronounced effect of the Kerr nonlinearity is the bistability of an optical resonant mode. If a resonant mode is excited in an optical resonator filled with a Kerr-nonlinear material, every point on the resonance profile sees a different refractive index, due to the enhanced intensity. This difference in the refractive index results in the shift of the resonant frequency. If the shift surpasses a certain threshold, the whole resonance profile appears to lean to one side, and a part of it becomes bistable, i.e. two stable amplitudes correspond to a given driving frequency \cite{Boyd2008}. This process is called `bistable transition'.

A microcavity filled with a Kerr-nonlinear material can exhibit the bistability and the bistable transition of a WGM. Up to now, the bistabilty of a WGM has been studied for various applications, such as optical switches or filters \cite{Kippenberg2002,Rodriguez2007}. In this work, however, we are going to focus on some other aspect of the Kerr effect in a microcavity, namely the relationship between dynamical tunneling and the cubic nonlinearity. As a solution of Maxwell's equations, a WGM has an intrinsic time-reversal symmetry, which manifests itself in two counter-propagating waves. Owing to the high $Q$ factor again, the degeneracy of a WGM can be lifted by a small perturbation, such as Rayleigh scattering \cite{Mazzei2007,Treussart1998}. The doublet of a WGM can be interpreted as a state resulting from tunneling between two distinct dynamical states, namely clockwise and counterclockwise propagating waves. Since these two coupled waves occupy the same spatial domain, we can find an analogy with dynamical tunneling \cite{dtunnel1, dtunnel2}, in contrast to a conventional tunneling between spatially separate domains. The bistable transition of such a WGM doublet has been already experimentally observed \cite{Treussart1998}. 

For a conventional resonant tunneling, it is well known that a cubic nonlinearity such as the Kerr nonlinearity can suppress the tunneling rate, because the nonlinear effect can break the symmetry when the system is unevenly excited. Since the cubic nonlinearity appears also in the governing equation of a bosonic many-body system described by mean-field approximation, the suppression of the conventional tunneling has been well studied with Bose-Einstein condensates \cite{BEC}. In recent years, some researches have been focused on the possibility to address the general issues of bosonic many-body systems such as the suppression of tunneling, using the Kerr nonlinearity \cite{kerrboso1, kerrboso2}. 

In this work, we extend the scope of investigation into the interplay between Kerr-nonlinearity and tunneling to the concept of dynamical tunneling. Since dynamical tunneling can occur between two spatially overlapping domains, the suppression of tunneling cannot be expected to occur for  dynamical tunneling in the same way as for conventional resonant tunneling. Even in the doublet of a WGM  two counter-propagating modes share the same spatial domain, hence the effect of the cubic nonlinearity on the tunneling is more subtle. With this motivation, we investigate how the Kerr nonlinearity affects the doublet of WGMs, using numerical and theoretical models. As a model system, we choose the `add-drop filter’\cite{Manolatou1999}, that comprises a microcavity with two wave guides coupled to the cavity, and analyze its transmission and resonant modes numerically and theoretically, varying the Kerr nonlinearity of the cavity. 

In the numerical simulation of the add-drop filter, it is observed that one directional whispering gallery mode gets dominant when the Kerr nonlinearity is increased. The reason behind this is mainly that the perturbation that lifts the degeneracy between the states also leads to uneven linewidths of the doublet, and they result in different bistable transitions. Using the theoretical model based on `coupled mode theory(CMT)’ \cite{Manolatou1999}, we analyze the details of the mode dynamics behind the phenomena.

In the next section, the add-drop filter will be first introduced, and the numerical simulation of its transmission and WGM doublets will be  presented. In Section \ref{seciii}, the effect of the nonlinearity on the transmission and the WGMs will be investigated. To analyze the properties of the mode dynamics, such as  the rotation and the stability of WGMs, a theoretical model will be built up on the basis of CMT \cite{Manolatou1999}. By analyzing the mode dynamics, the physical background behind the observed phenomena will be clarified.
\section{Add-Drop filter and CMT modeling}\label{secii}
\begin{figure}
\begin{center}
\includegraphics[width=.85\columnwidth]{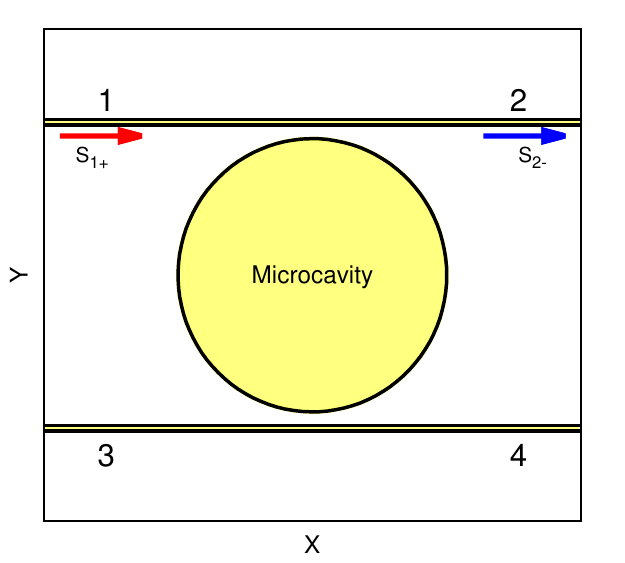}
\caption{(color online) Structure of the add-drop filter: a circular microcavity with radius $R$ is coupled to two wave guides. The four branches given by the waveguides are indexed by $i=1,2,3$ and $4$. The coupling is through overlapped evanescent fields. The width of the waveguides is $0.04R$ and the closest gap between the cavity and the waveguides is $0.1R$. The transmission from the channel $1$ to the channel $2$ are numerically computed by the ratio of  the outgoing wave $S_{2-}$ to the incoming wave $S_{1+}$.}\label{fig1}
\end{center}
\end{figure}

The add-drop filter consists of a microcavity and two optical fibers side-coupled to the cavity through an evanescent field. The two wave guides provide effectively four channels. In Fig.~\ref{fig1}, the four channels are indexed by $i = 1,2,3$ and $4$. In this work, the field in the lower waveguide is not analyzed. However, we keep it in the scheme for two reasons. First, the lower waveguide makes the geometry of the system two-fold symmetric that helps lift the degeneracy of WGMs more easily. Second, if the waveguide which has the incoming wave is not experimentally available for the measurement of the outgoing wave, the mode dynamics presented in this work can be investigated by measuring the outgoing wave in the lower waveguide.

In this add-drop filter, the shortest distance between the cavity boundary and the wave guide is set equal to $0.1R$, where $R$ is the radius of the cavity. The refractive index of the cavity and the wave guide is chosen to be $n_{0}=2.0$. In order to induce a strong backscattering of evanescent field and to thereby lift the degeneracy of a WGM more easily, very narrow wave guides with the width of $0.04R$ are put at the upper and the lower sides of the cavity. To investigate the transmission, a continuous wave source with frequency $\Omega$ is set at the channel $1$, which provides the incoming wave $S_{1+}$. In the numerical analysis, the amplitude of the source $S_{1+}$ is set equal to $10$, and the wave arriving in the channel $2$ $S_{2-}$ is computed by means of `Finite Difference Time Domain(FDTD)' method \cite{fdtd, Oskooi2010}. 
\begin{figure}
\begin{center}
\includegraphics[width=\columnwidth]{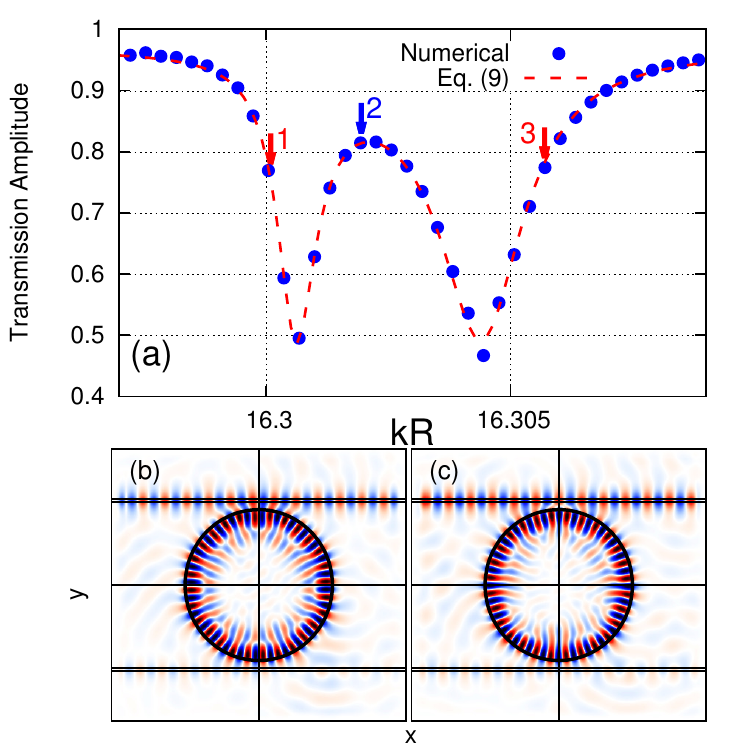}
\caption{(color online) (a) Transmission amplitude at channel 2 without Kerr nonlinearity. Due to the scattering on the waveguides and the high-$Q$ factor of the resonance, the transmission dip is split. The excited waves in the cavity on the left-hand side and the right-hand of the split dips(denoted by $1$ and $3$) exhibit clockwise rotation, whereas the wave between the split dips(denoted by $2$) shows counterclockwise rotation. The result of the comparison with Eq.~(\ref{coupletrans}) is superimposed as a dotted curve. Each dip exhibits different parity((b) odd (c) even) with respect to the symmetry axes of the system.}\label{fig2}
\end{center}
\end{figure}

If a mode with an extremely high cavity-$Q$ factor is excited in the microcavity, the coupling via scattering can lift the intrinsic degeneracy of the doublet state and split the modes \cite{Kippenberg2002,Mazzei2007}. The scattering on the two wave guides attached to the cavity is adjusted to be strong enough to lift the degeneracy. By the FDTD simulation, such a coupling is observed. Figure~\ref{fig2} shows that a whispering gallery mode with the radial mode number $l=1$ and the azimuthal mode number $m=28$ is coupled to the wave guide. As the cavity modes have extremely high $Q$ factors, the transmission curve exhibits two split dips, and each dip corresponds to a different parity with respect to reflectional symmetric axis. Figures~\ref{fig2}(b) and (c) show the two modes corresponding to the left and the right dips in the transmission curve in Fig.~\ref{fig2}(a), which display odd and even symmetry, respectively. 

Another noteworthy feature can be found in the rotation of the excited wave, which depends on the relative position to the split peaks.  As the incoming wave runs from the left-hand to the right-hand side of the upper waveguide in Fig.~\ref{fig1}, it is intuitively expected that the wave in the cavity would be launched on the upper side of the cavity parallel to the incoming wave, thus the excited field in the cavity would rotate in the clockwise direction. However, they exhibit the counterclockwise field rotation in between the split dips, contrary to the intuitive expectation, whereas on the right-hand and the left-hand side of the split dips they show the more intuitive clockwise rotation \cite{Mazzei2007}. 

The splitting of the WGMs is analyzed by using `coupled mode theory'. For the sake of intuitive modeling, two counter-rotating modes with the same resonant frequency $\omega$ are initially adopted as a basis, and the following equations are derived:
\begin{align}
\frac{d a_{c}}{d t} &= (i(\omega+g)-\gamma - \frac{1}{2}\Gamma) a_{c} + (ig - \frac{1}{2}\Gamma) a_{cc} + \eta_{c} S_{1+} \label{doublecmt1} \\
\frac{d a_{cc}}{d t} &= (i(\omega + g) - \gamma - \frac{1}{2}\Gamma) a_{cc} + (ig - \frac{1}{2}\Gamma) a_{c} + \eta_{cc} S_{1+}.
\label{doublecmt2}
\end{align}
Here the subscript $c$ and $cc$ stand for clockwise and counterclockwise modes, $g$ is the coupling parameter given by the scattering. $\gamma$ is the characteristic attenuation of the WGM given the mode configuration \cite{Nockel1997, Shim2013}, and $\Gamma$ is the additional attenuation induced by the scattering. Although the source is put only in channel $1$, it is necessary to assume the driving $\eta_{cc}$, because the counterclockwise mode is also driven by scattering on the waveguides.

Since the modes associated with the two transmission dips are two standing wave modes, the basis is transformed to
\begin{equation}
\boldsymbol{\psi}_{+} = \dfrac{\boldsymbol{\psi}_{c}+\boldsymbol{\psi}_{cc}}{2}~~~\text{and}~~~\boldsymbol{\psi}_{-} = \dfrac{\boldsymbol{\psi}_{c}-\boldsymbol{\psi}_{cc}}{2i}.
\label{standing}
\end{equation}
For the circular microcavity, these modes are represented by the product of a Bessel function $J_m$ and a $\sin$ or $\cos$ function according to
\begin{equation}
\begin{aligned}
\boldsymbol{\psi_+}( r, \theta ) &= J_m(nkr) \cos (m\phi)\\
\boldsymbol{\psi_-}( r, \theta ) &= J_m(nkr) \sin (m\phi),
\label{standingbasis}
\end{aligned}
\end{equation}
where $n$ is the refractive index of the cavity and the wave guide, and $k$ is the resonant wave number of the microcavity with the azimuthal mode number $m$. The resonant wave number is determined by solving Maxwell equations with the given microcavity boundary and mode numbers. Corresponding to the change of basis, the mode equations in Eqs.~(\ref{doublecmt1}) and (\ref{doublecmt2}) are also rearranged in the form
\begin{align}
\frac{d a_{+}}{d t} &= (i(\omega + 2 g) - \gamma - \Gamma) a_{+}  +  \eta_{+} S_{1+}
\label{doublecmt3-0}\\
\frac{d a_{-}}{d t} &= (i\omega-\gamma) a_{-}  + \eta_{-} S_{1+},
\label{doublecmt3}
\end{align}
where the driving strengths are given by 
\begin{equation}
\eta_{+} = \dfrac{\eta_{c} + \eta_{cc}}{2} ~~~\mbox{and}~~~ \eta_{-} = \dfrac{\eta_{c} - \eta_{cc}}{2i}.
\end{equation}
Then, the transmission amplitude at the channel $2$ can be formulated as
\begin{align}
t &= 1 - \dfrac{\eta^*_{+} a_{+}}{S_{1+}} - \dfrac{\eta^*_{-} a_{-}}{S_{1+}} \label{trans1}\\
&= 1 - \dfrac{|\eta_+|^2}{i(\omega+2g-\Omega) - \gamma  -\Gamma} - \dfrac{ |\eta_-|^2}{i(\omega -\Omega)- \gamma}. \label{coupletrans}
\end{align}
\linebreak
Because some parameters in Eq.~(\ref{coupletrans}) cannot be analytically evaluated, the values of the couplings and attenuations are obtained by taking them as free parameters and comparing Eq.~(\ref{coupletrans}) with the numerical result from the FDTD simulation in Fig.~\ref{fig2}. This yields
\begin{align}
|\eta_{+}|^2&= 5.043 \times 10^{-5}, \nonumber \\
|\eta_{-}|^2&= 1.159 \times 10^{-4}, \nonumber \\
\gamma&= 1.118 \times 10^{-4}, \nonumber \\
\Gamma&= 1.135 \times 10^{-4}.
\label{fittingresult}
\end{align}
%
To obtain these values, the contribution of the neighboring modes to the transmission is carefully considered. In the spectrum of a circular microcavity, the $(m, l) = (20,3)$- and the $(24,2)$-modes ($m$: azimuthal mode number, $l$: radial mode number) are standing on the left-hand and the right-hand sides of the doublet, respectively. Their contributions to the transmission are extracted prior to the comparison. 

\section{Kerr nonlinearity in Numerical Computation and CMT modeling}\label{seciii}
In a Kerr medium, the electric field and the electric displacement are related by the nonlinear equation \cite{Boyd2008}
\begin{equation}
\label{kerr1}
\mathbf{D} = \epsilon_0 (1+\chi^{(0)}) \mathbf{E}+  \epsilon_0 \chi^{(3)}  \mathbf{E}^3.
\end{equation}
By substituting a monochromatic electromagnetic wave $\mathbf{E} =\mathbf{E_{\omega}} \exp (i \omega t)$ into Eq.~(\ref{kerr1}), the third-order nonlinear term in Eq.~(\ref{kerr1}) leads to
\begin{align}
(\text{Re}[\mathbf{E}_\omega e^{-i\omega t}])^3 &= \nonumber \\ (1/4) \text{Re}[\mathbf{E}^3_\omega &e^{-3i\omega t} ] +(3/4) |\mathbf{E_\omega}|^2 \text{Re}[\mathbf{E}_\omega e^{-i\omega t}].
\end{align}
By neglecting the third-harmonic term, the effective refractive index can be written in the well-known form,
%
\begin{equation} \label{kerrdef}
n = n_0 + \dfrac{3 \chi^{(3)}}{4 n^2_0} |\mathbf{E}|^2.
\end{equation}
\begin{figure}
\begin{center}
\includegraphics[width=\columnwidth]{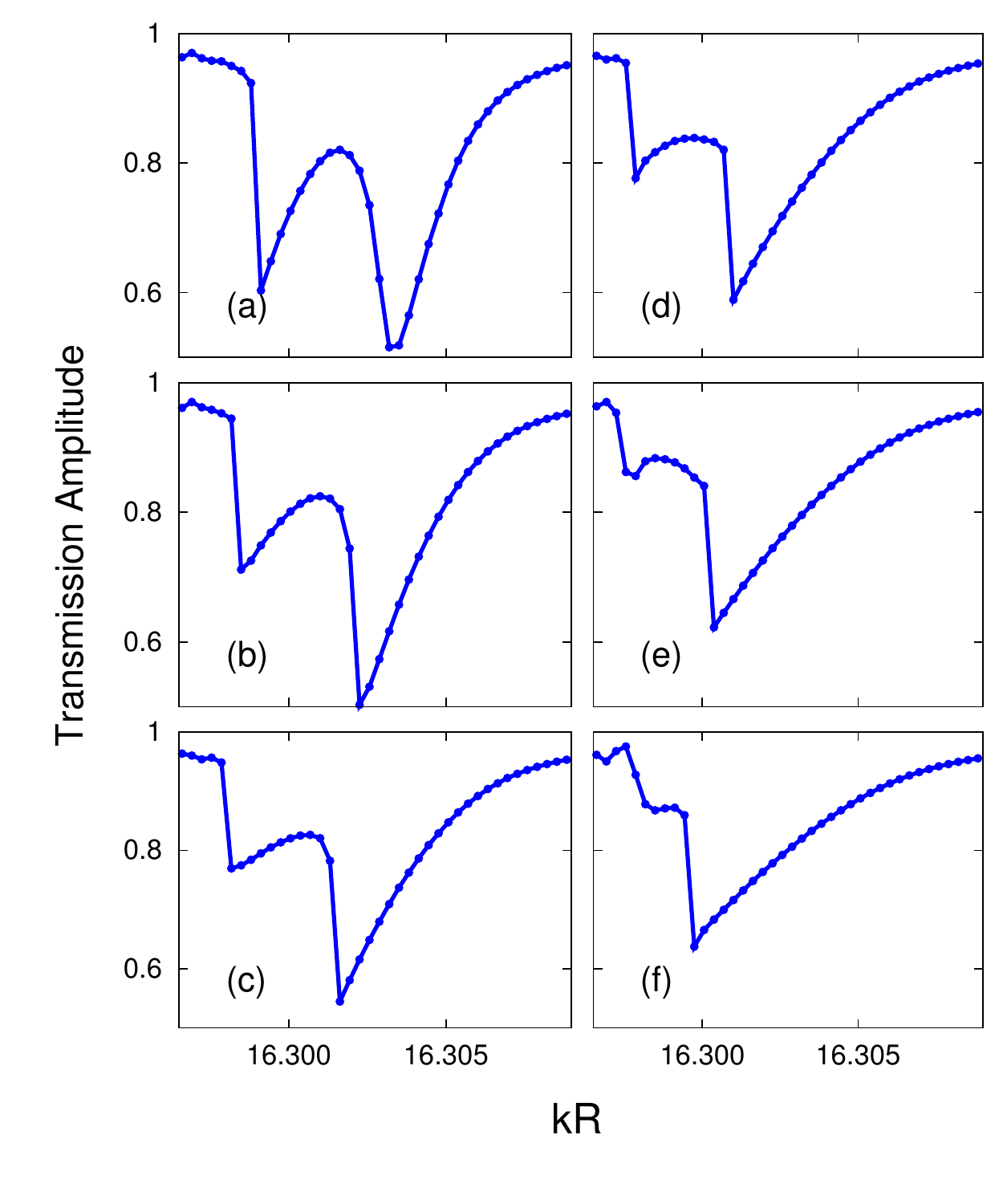}
\caption{(color online) FDTD simulation of the transmission amplitude associated with the split whispering gallery modes. (a) $\chi^{(3)} = 1.0 \times 10^{-7}$, (b) $2.0 \times 10^{-7}$, (c) $3.0 \times 10^{-7}$, (d) $4.0 \times 10^{-7}$, (e) $5.0 \times 10^{-7}$, (f) $6.0 \times 10^{-7}$. The intensity of the whispering gallery modes lies in the range of $|\mathbf{E}|^2 = 2 \sim 6 \times 10^{4}$.}\label{nldoublet}
\end{center}
\end{figure}
The effect of the Kerr nonlinearity on the split whispering gallery modes is first numerically investigated. Figure~\ref{nldoublet} presents the FDTD computation of the transmission amplitude through the add-drop filter and its evolution with the increase of $\chi^{(3)}$. As the nonlinearity increases, the resonance profiles in the transmission curves lean more toward the left-hand side. When the nonlinearity exceeds a certain value, it is noticed that an abrupt upward shift occurs in both of the transmission dips. This is an evidence of the bistable transition, because unstable mode dynamics is involved in the parts where the abrupt shifts occur.

The bistable transition progresses gradually with the increase of $\chi^{(3)}$. The remarkable feature in this transition is the difference in the bistable transition of the two dips. Namely, the broader a dip is, the slower it goes bistable with increase of the nonlinearity. For instance, at $\chi^{(3)} = 1.0 \times 10^{-7}$ the left-hand side dip shows the bistability, whereas the right-hand side dip does not yet. Obviously, this difference results from the uneven linewidth of the doublet. As seen in the previous section, the two dips of the doublet states have different damping rates in the linear regime, that are given by $\gamma$ and $\Gamma + \gamma$, respectively for the right- and the left-hand side dips. This uneven transition to bistability leads to two uneven depths of the dips in the transmission profile, as shown in Fig.~\ref{nldoublet}. Another noticeable observation in this computation is that the clockwise rotation of the field gets dominant when the value of $\chi^{(3)}$ is sufficiently large, such as $5 \times 10^{-7}$. As the transmission dips lean to the left-hand side, the spectral region between the two dips which shows the counterclockwise field rotation shrinks, and the transmission dip on the left-hand side vanishes. In contrast, the right-hand side curve of the dip remains and keeps the clockwise field rotation as it has in the linear optical regime.
\begin{figure}
\begin{center}
\includegraphics[width=\columnwidth]{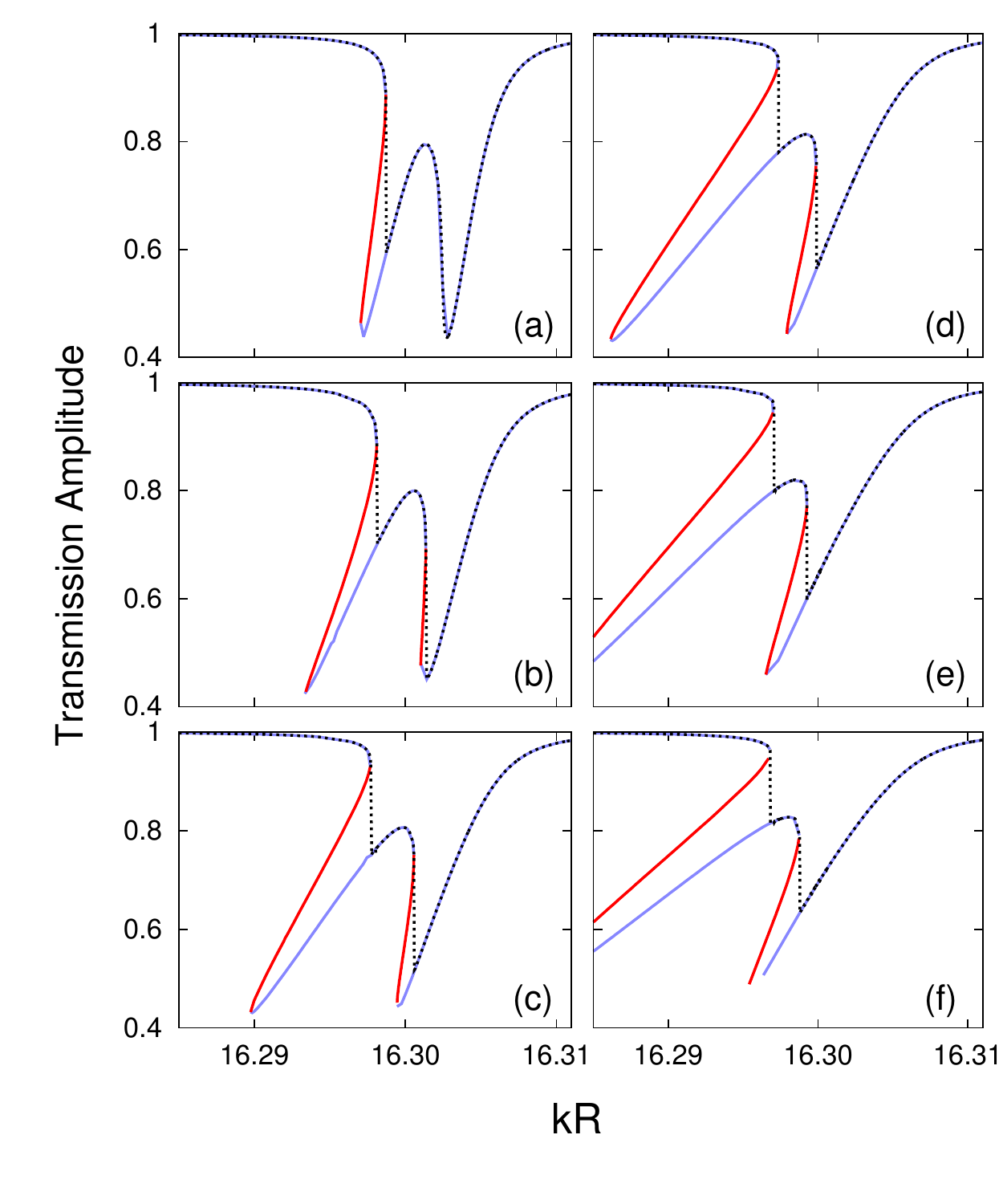}
\caption{(color online) Results of the CMT analysis for the split transmission under $\chi^{(3)}$ nonlinearity. With increasing nonlinearity, stationary states are identified and plotted. Stable and unstable branches of the profile are denoted by blue and red colors, respectively. (a)  $\chi^{(3)} = 1.0 \times 10^{-7}$, (b) $2.0 \times 10^{-7}$, (c) $3.0 \times 10^{-7}$, (d) $4.0 \times 10^{-7}$, (e) $5.0 \times 10^{-7}$, (f) $6.0 \times 10^{-7}$. The superimposed black dotted lines are connecting the branches that are most likely to be observed in numerics and in an experiment. They show a good agreement with Fig.~\ref{nldoublet}.}\label{nldoublet2}
\end{center}
\end{figure}

Using the CMT model, this numerical observation can be reconstructed, and the underlying mode dynamics behind the observation can be addressed more clearly. To apply the CMT model to the system, the Kerr-nonlinear effect can be incorporated in the model by treating it as a perturbation. For other values necessary for the model, such as attenuations and resonant frequencies, those in Eq.~(\ref{fittingresult}) are used. The electric field in the cavity has to be first represented with the basis functions, $\psi_{+}$ and $\psi_{-}$, as follows:
\begin{equation}
\label{combination}
\mathbf{E} = a_{+}(t) \boldsymbol{\psi_{+}}(\mathbf{r}) + i a_{-}(t) \boldsymbol{\psi_{-}}(\mathbf{r}).
\end{equation}
Then, the field intensity formed in the cavity is proportional to
\begin{equation}
|\mathbf{E}|^2 = (a^{*}_{+} \boldsymbol{\psi_{+}} - i a^{*}_{-} \boldsymbol{\psi_{-}})\cdot(a_{+} \boldsymbol{\psi_{+}} + i a_{-} \boldsymbol{\psi_{-}}).
\end{equation}
By applying the perturbation approach to Eq.~(\ref{kerrdef}) \cite{Grigoriev2011}, the mode equations in Eqs.~(\ref{doublecmt3-0}) and (\ref{doublecmt3}) are modified to
\begin{align}
\frac{d a_{+}}{d t} &= (i(\omega + 2 g) - \gamma - \Gamma) a_{+} + i\omega \kappa_{++} |a_{+}|^2 a_{+} \nonumber \\
&+ i\omega \kappa_{+-}(2a_{+}|a_{-}|^2 -a^{*}_{+} a^2_{-})+ \eta_{+} S_{1+}\label{doublecmt4}\\
\frac{d a_{-}}{d t} &= (i\omega-\gamma ) a_{-}  + i\omega \kappa_{--} |a_{-}|^2 a_{-} \nonumber \\
&+ i \omega \kappa_{+-}(2a_{-}|a_{+}|^2 -a^{*}_{-} a^2_{+})+\eta_{-} S_{1+},
\label{doublecmt5}
\end{align}
where $\kappa_{++}$, $\kappa_{+-}$ and $\kappa_{--}$ are given by 
\begin{equation}
\kappa_{pq} = \dfrac{3\chi^{(3)}}{4n^2} \dfrac{\int |\psi_{p}|^2 |\psi_{q}|^2dV}{\int |\psi_{+}|^2 + |\psi_{-}|^2 dV}, ~\text{where}~~~~
p, ~q = +~ \text{or}~ -.
\end{equation}
If the standing wave modes in Eq.~(\ref{standingbasis}) are substituted into the above equation, the $\kappa$'s are related by
\begin{equation}
\kappa_{++} = \kappa_{--} = 3 \kappa_{+-}.
\end{equation}
\linebreak
In order to search for stationary states of the mode equations, the following ans\"atze are first introduced:
\begin{align}
\label{ansatz0-1}
a_{+}&= A_{+} e^{i (\Omega t + \phi_{+})},\\
a_{-}&=A_{-} e^{i (\Omega t + \phi_{-})}. \label{ansatz0-2}
\end{align}
By substituting Eqs.~(\ref{ansatz0-1}) and (\ref{ansatz0-2}) into Eqs.~(\ref{doublecmt4}) and (\ref{doublecmt5}), the mode equations are reduced to
\begin{align}
i \Omega A_{+} &= (i(\omega +2g) - \gamma - \Gamma)A_{+} + i3\omega \kappa A_{+}^3 \nonumber\\
&+ i\omega \kappa ( 2 A_{+} A_{-}^2 - A_{+} A_{-}^2 e^{-i2(\phi_{+} - \phi_{-})}) + K_{+} e^{-i \phi_{+}} \label{stationary1},\\
i \Omega A_{-} &= (i\omega - \gamma )A_{-} + i3\omega \kappa A_{-}^3 \nonumber\\
&+ i\omega \kappa ( 2 A_{-} A_{+}^2 - A_{-} A_{+}^2 e^{i2(\phi_{+} - \phi_{-})}) - i K_{-} e^{-i \phi_{-}} \label{stationary2},
\end{align}
where $K_\pm$ are given by $|\eta_\pm|K$ and the amplitude $K$ of the source is set equal to $10$, as the source amplitude in the numerical computation. 

By numerically optimizing Eqs.~(\ref{stationary1}) and (\ref{stationary2}), stationary states are identified at a given nonlinearity $\kappa$ and a driving frequency. To facilitate the comparison with the full numerical solution, the obtained stationary states are converted to transmission amplitude by using Eq.~(\ref{trans1}). The result of the computation reveals the underlying bistability profile of the full numerical solution. Figure~\ref{nldoublet2} shows a bistable transition in the transmission profiles, following the increase of nonlinearity.
\begin{figure}
\begin{center}
\includegraphics[width=\columnwidth]{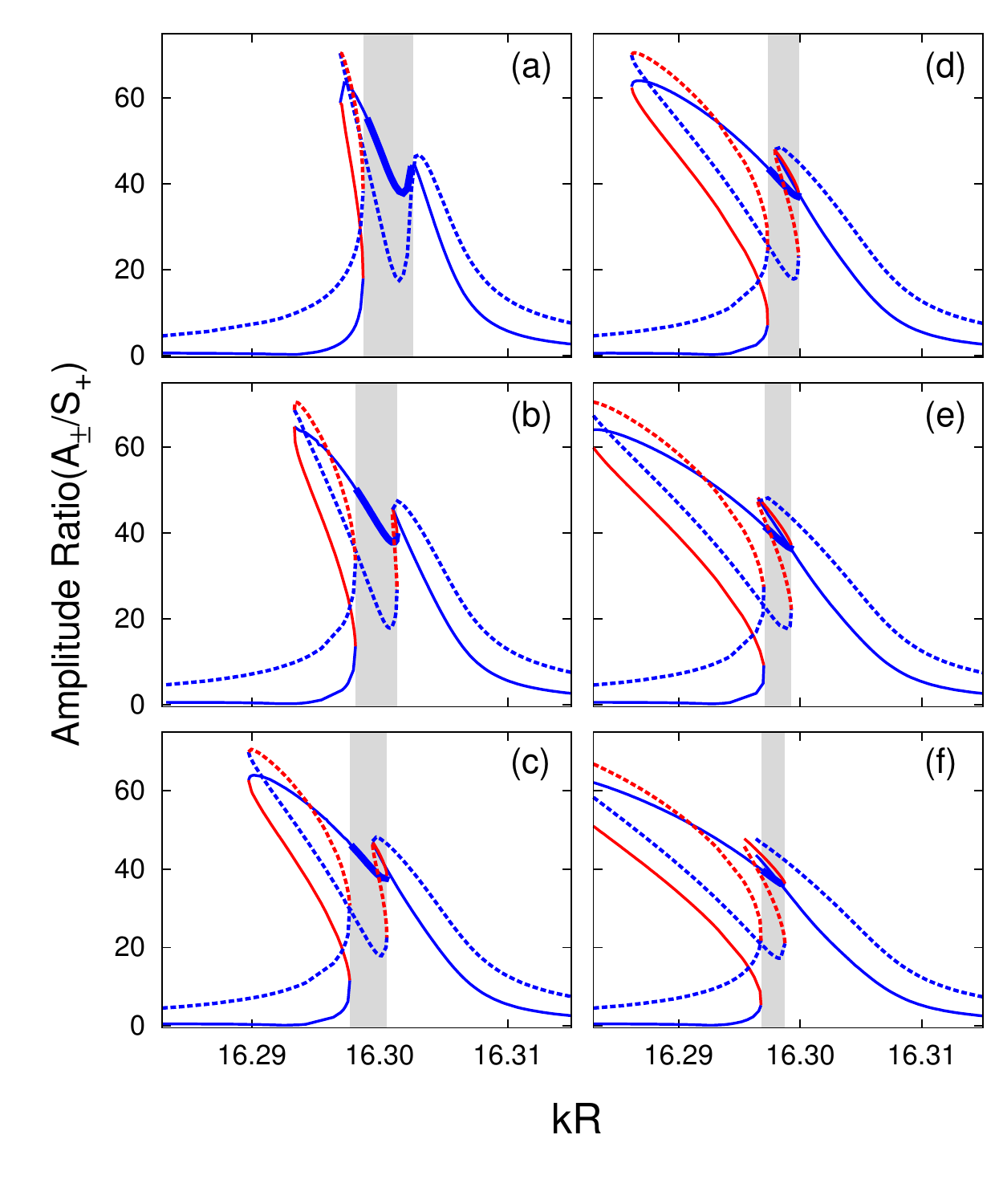}
\caption{(color online) Relative amplitudes of  clockwise-(dotted) and counterclockwise(solid) rotating modes with respect to the driving amplitude $S_+$. The stability of each branch is denoted by the same color code as in Fig.~\ref{nldoublet2}(stable: blue, unstable:red). (a) $\chi^{(3)} = 1.0 \times 10^{-7}$, (b) $2.0 \times 10^{-7}$, (c) $3.0 \times 10^{-7}$, (d) $4.0 \times 10^{-7}$, (e) $5.0 \times 10^{-7}$, (f) $6.0 \times 10^{-7}$. The range where the counterclockwise rotating mode is dominant is shaded by grey colour, and the corresponding amplitude that is likely dominant in this range is denoted by a thick line.}\label{nldoublet-modes}
\end{center}
\end{figure}
As mentioned in the introduction, when the $\chi^{(3)}$ nonlinearity exceeds a certain threshold, three stationary states correspond to one frequency in a bistable range. In this case, the two states with the highest and the lowest amplitudes have stable mode dynamics around themselves, whereas one in the middle has unstable dynamics. 

In order to measure the stability of stationary states, another ans\"atze are introduced for $a_{+}$ and $a_{-}$:
\begin{equation}
a_{+} = (\alpha_{+} + i \beta_{+}) e^{i \Omega t} ~~\mbox{and}~~a_{-} = (\alpha_{-} + i \beta_{-}) e^{i \Omega t},
\end{equation}
where the real and the imaginary parts of the complex amplitudes $\alpha$ and $\beta$ are time-dependent unlike the ans\"atze in Eqs.~(\ref{ansatz0-1}) and (\ref{ansatz0-2}). 

By substituting these ans\"atze into Eqs.~(\ref{doublecmt4}) and (\ref{doublecmt5}) and by separating the real- and the imaginary parts, the following four coupled equations are derived.
\begin{align}
\dfrac{d \alpha_{+}}{dt} = &-(\gamma+\Gamma)\alpha_{+}-(\omega+2g - \Omega)\beta_{+}- 3\omega \kappa (\alpha_{+}^2+\beta_{+}^2) \beta_{+} \nonumber\\
-& 2 \omega \kappa (\alpha_{-}^2 +\beta_{-}^2) \beta_{+} + \kappa \omega (2\alpha_{-} \beta_{-} \alpha_{+} - \alpha_{-}^2 \beta_{+} + \beta_{-}^2 \beta_{+}) \nonumber\\
+&K_{+},\nonumber \\
\dfrac{d \beta_{+}}{dt} = &(\omega+2g - \Omega)\alpha_{+} - (\gamma + \Gamma) \beta_{+} + 3 \omega \kappa (\alpha_{+}^2 + \beta_{+}^2) \alpha_{+} \nonumber\\
+&2 \omega \kappa (\alpha_{-}^2+\beta_{-}^2) \alpha_{+} + \kappa \omega (\alpha_{+} \beta_{-}^2 -\alpha_{+} \alpha_{-}^2 - 2\alpha_{-} \beta_{-} \beta_{+}), \nonumber \\
\dfrac{d \alpha_{-}}{dt} =& -\gamma \alpha_{-}-(\omega - \Omega)\beta_{-} - 3\omega \kappa (\alpha_{-}^2+\beta_{-}^2) \beta_{-} \nonumber\\
- &2 \omega \kappa (\alpha_{+}^2 +\beta_{+}^2) \beta_{-} + \kappa \omega (2\alpha_{+} \beta_{+} \alpha_{-} - \alpha_{+}^2 \beta_{-} + \beta_{+}^2 \beta_{-}), \nonumber \\
\dfrac{d \beta_{-}}{dt}  = & (\omega - \Omega)\alpha_{-} - \gamma \beta_{-} + 3 \omega \kappa (\alpha_{-}^2 + \beta_{-}^2) \alpha_{-} \nonumber\\
+&2 \omega \kappa (\alpha_{+}^2+\beta_{+}^2) \alpha_{-} + \kappa \omega (\alpha_{-} \beta_{+}^2 -\alpha_{-} \alpha_{+}^2 - 2\alpha_{+} \beta_{+} \beta_{-}) \nonumber\\
-&K_{-}.
\end{align}
The stability at given amplitudes can be inferred from the Jacobian matrix of the above four equations. If the eigenvalues of the Jacobian matrix   are all negative, the associated stationary state has a stable mode dynamics around it, while an unstable dynamics is inferred otherwise \cite{Ott2002}. The result of the stability computation is encoded on the resonance profile with a color code. In Fig.~\ref{nldoublet2}, stable and unstable branches are denoted by blue and red colors, respectively. 

If the stability of a stable branche is quantified by the smallest absolute value of the eigenvalues of the Jacobian matrix, the higher stable branch in the transmission curve has a larger stability that the lower one. Therefore, the upper branch is more likely to be numerically converged and experimentally observed. In Fig.~\ref{nldoublet2}, the curves connecting upper stable branches are superimposed as black dotted lines. These curves show a good agreement with numerical results in Fig.~\ref{nldoublet}. 

In this computation, the difference in the bistable transition of the two dips is also observed. As seen in Fig.~\ref{nldoublet}, the right-hand side dip in Fig.~\ref{nldoublet2} is still completely monostable at $\chi^{(3)} = 1\times 10^{-7} $, whereas the left-hand side dip already gets into the bistable regime. This is apparently caused by the uneven damping of the doublet. Since the two dips have different dampings, namely $\gamma$ and $\gamma + \Gamma$ in the linear regime, the right-hand side dip with higher damping shows slower bistable transition. In the end, the right-hand side dip takes most part of the dotted curve in Fig.~\ref{nldoublet2}, whereas the left-hand side dip is strongly suppressed when it goes to the bistable regime. At the same time, the spectral region between the two dips which shows the counterclockwise field rotation, shrinks. In the bistable transition, each part of the transmission dip keep the field rotation. Hence, WGMs dominantly show a clockwise rotation over the transmission dips when the system gets into the strongly bistable regime.

To render the mode dynamics behind this phenomenon clearer, the associated rotating amplitudes are reconstructed by CMT modelling, again. For this purpose, the basis functions are transformed back to the rotational basis by reversing Eq.~(\ref{standing}). Figure~\ref{nldoublet-modes} shows the result of the transformation. In Fig.~\ref{nldoublet-modes}, the amplitude of the clockwise rotating mode is displayed by a dotted curve, and that of the counterclockwise rotating one is represented by a solid curve. In the bistable transition, one of these curves exhibits a twist of the peak, whereas the other one shows a leaning peak to the left-hand side. By comparing the amplitude of the clockwise- and the counterclockwise modes, the part of the curve where the counterclockwise rotation is dominant is identified and shaded with grey colour in Fig.~\ref{nldoublet-modes}. In this comparison, the amplitude in the lower stable branch is chosen in the bistable region, because the lower branch has the higher stability. The figure shows an apparent tendency, namely that the region of the counterclockwise rotation shrinks with the increase of nonlinearity. In other words, the clockwise becomes dominant over the resonant profile, as is also numerically observed. 

\section{Conclusion}
The effect of a Kerr nonlinearity on coupled whispering gallery modes is investigated numerically and theoretically. In this study, an optical add-drop filter which consists of a microcavity and two side-coupled wave guides is chosen as a model system. When a whispering gallery mode with extremely high $Q$ factor is induced in the microcavity, the intrinsic degeneracy of the whispering gallery mode can be lifted by scattering of the evanescent field, thus forming a doublet state. Correspondingly, the transmission curve also exhibits two split downward peaks. The evolution of these split peaks with increasing nonlinearity is first numerically investigated by using FDTD algorithm. In this computation, the transmission shows a typical bistable transition under the influence of the cubic nonlinearity. However, the two peaks of the doublet show different transition to bistability, since the two peaks have a different damping. The different damping is caused by the difference in scattering rates, depending on the mode parity. As a result, one side of the peak with higher damping takes a large part of the transmission curve, as soon as the bistability is induced. Thus, the rotation of field that is mainly observed on this side of the peak gets dominant.

These numerically observed phenomena are addressed by a Coupled-Mode-Theory(CMT) model. Using the perturbation approach, we incorporate the nonlinear optical effect in the CMT model, and obtain the dynamics of modes in microcavity. This theoretical modelling shows a quantitative consistency with the numerical observations, and reveals how the one-directional rotation of field becomes dominant with increasing nonlinearity.

Using the theoretical analysis developed in this work, we anticipate that further issues in general transport phenomena under cubic nonlinear effects can be treated
. 
\\
\begin{acknowledgements}
This work is financially supported by Deutsche Forschungsgemeinschaft (DFG) within the framework of Forschergruppe FOR760.
\end{acknowledgements}

\end{document}